\documentclass[11pt,preprint]{aastex}

\usepackage{graphicx}

\newcommand{\xmm}{{\em XMM-Newton}}
\begin{document}                                                                     

\title{XMM-Newton Observation of the $\alpha$ Persei Cluster }


\author{Ignazio Pillitteri }  
\affil{Smithsonian Astrophysical Observatory,  MS 6,   
 60 Garden St., Cambridge, MA 02138; email ipillitteri@cfa.harvard.edu}

\author{Nancy Remage Evans$^1$ }
\affil{Smithsonian Astrophysical Observatory,  MS 4,   
 60 Garden St., Cambridge, MA 02138; email: nevans@cfa.harvard.edu}

\author{Scott J. Wolk}  
\affil{Smithsonian Astrophysical Observatory,  MS 70,   
 60 Garden St., Cambridge, MA 02138}

\author{Megan Bruck Syal}  
\affil{Department of Geological Sciences, Brown University, 
Box 1846, Providence, RI 02912 }



\begin{abstract}
We report on the analysis of an archival observation of part of the
$\alpha$ Persei cluster
obtained with \xmm.  
We detected 102 X-ray sources in the band 0.3-8.0 keV, of which  
39 of them are   associated with the cluster as evidenced by
appropriate magnitudes and colors from  
 2MASS photometry. We extend the 
X-ray Luminosity Distribution (XLD) for
M dwarfs,  to add to the XLD found for hotter dwarfs from spatially extensive surveys of
the whole cluster by ROSAT. Some of the hotter stars are identified as
 a background, possible slightly older
group of stars at a distance of approximately 500 pc.
\end{abstract}



\footnote[1]{Corresponding Author}

\section{Introduction}
Open clusters have been a keystone in understanding stellar evolution
because they contain stars  at the same
distance with similar reddening
formed at the same time with the same chemical composition,
at least to a first approximation.   Their members  display a range of
masses, temperatures, luminosities, rotation rates, and multiplicity
which can be explored.  
Comparison of cluster morphology provides an age sequence  as a
context for the evolution of stars. The well-known
spin-down of low mass stars with age due to magnetic braking provides
a good example of insight from clusters.  This slowing of  rotation results in
the decrease of coronal X-rays due to their connection with stellar
dynamos. Good summaries of  the decrease in stellar activity 
as stars age for a range of masses are provided  by Favata 
and Micela (2003) and G\"udel (2004).  

$\alpha$ Per is a young open
cluster, found to be 50 Myr old from upper main sequence turnoff morphology
(Meynet, Mermilliod, and Maeder, 1993).  More recently, Stauffer, et
al. (1999) have found an age of 90 Myr from the low mass lithium
depletion boundary.  Although there is some dispersion in the exact
calibration of the age of the cluster, the sequence of age (increasing
from the Orion Nebula Cluster through the $\alpha$ Per cluster through
the Pleiades) is generally agreed.  Thus, studies find a range in age
from 50 to 90 Myr.  We will use the shorthand ``50 Myr'' for the age
of the cluster.

This age makes it an 
excellent comparison for Cepheids and their companions.  Indeed, $\alpha$ Per
itself is a yellow supergiant, with similar parameters to  Cepheids, 
except for its location at a temperature
outside the instability strip.  As an example of usage of the cluster,
we have observed a number of Cepheids with the Hubble
Space Telescope Wide Field Camera 3  to identify a
population of resolved low-mass stars close to Cepheids which are
probable physical companions.  X-ray observations showing an
activity  level comparable to that of $\alpha$ Per dwarfs 
would confirm that they are young
companions rather than chance alignments with old field stars 
(e.g. Evans, et al. 2012).
Another interesting aspect of a cluster of this age is that it is the
period when young planets have just finished forming, and thus we gain
insight into the X-ray environment during the early formation of
atmospheres. 

Because the cluster is nearby, it covers a 
wide area in the sky. It was observed with a raster of pointings by {\it ROSAT}
(Randich, et al. 1996).  Essentially all the late F, G, and K members 
from the membership  studies of Prosser (1992) were detected.
Three deeper pointed {\it ROSAT}
observations (22-25 ks) were subsequently made covering part of the cluster
(Prosser, et al. 1996).  Two additional studies were made using  near IR 
observations to try to identify  counterparts of {\it ROSAT}
sources (Prosser and Randich 1998; Prosser, Randich, and Simon 1998, PRS below).  
In the second of 
these,  the authors identified a number of G and K stars with lower X-ray 
luminosity than expected for cluster members, which they termed ``bad L$_X$ stars''. 
Finally, a deeper (60 ksec) exposure of a small part of the cluster was made with XMM, 
which was described briefly by Pallavicini, Franciosini, and Randich (2004).  
This is the observation which we discuss here,
which allows us to investigate fainter sources as well as 
the spectral properties of the sources.  
In addition to being relatively nearby (170 pc; Randich, et
al. 1996), the $\alpha$ Per cluster is also lightly reddened 
(E(B-V) = 0.09 mag; Meynet, et al. 1993), making the data interpretation
relatively robust.  

The supergiant $\alpha$ Per itself was detected in the ROSAT observations (Prosser, 
et al. 1996).  However, recently Ayres (2011) has suggested that there is 
evidence that the X-rays might actually 
be produced by a  low-mass X-ray active companion.

In this study we add the results of the deeper XMM image to the existing literature
on the cluster.  Specifically, in the sections that follow,  we discuss the observations, 
the source detection and near IR matches, the source parameters (luminosity and spectra),
the X-ray luminosity distribution (XLD), light curves, and the
results.  Of particular importance in deriving the cluster parameters
 (Discussion section, Section 4) is a grouping of stars likely to be a
cluster of young stars behind the $\alpha$ Per cluster.

\section{Observations and Data Analysis}
A fraction of the $\alpha$ Per cluster was observed by 
XMM-Newton as part of the  Mission Scientist Guaranteed Time 
(Pallavicini, et al 2004). 
A 60 ks observation was obtained in Sept 5th 2000-09-05 using EPIC MOS and PN cameras
on board \xmm\/ with a pointing at  R.A.: $3^h26^m16^s$ and Dec.: $48^d50^m29^s$.
Fig. 1 shows the composite PN, MOS~1 and MOS~2 image of the EPIC field within the
$\alpha$ Per cluster. 

\begin{figure}
 \includegraphics[width=\columnwidth]{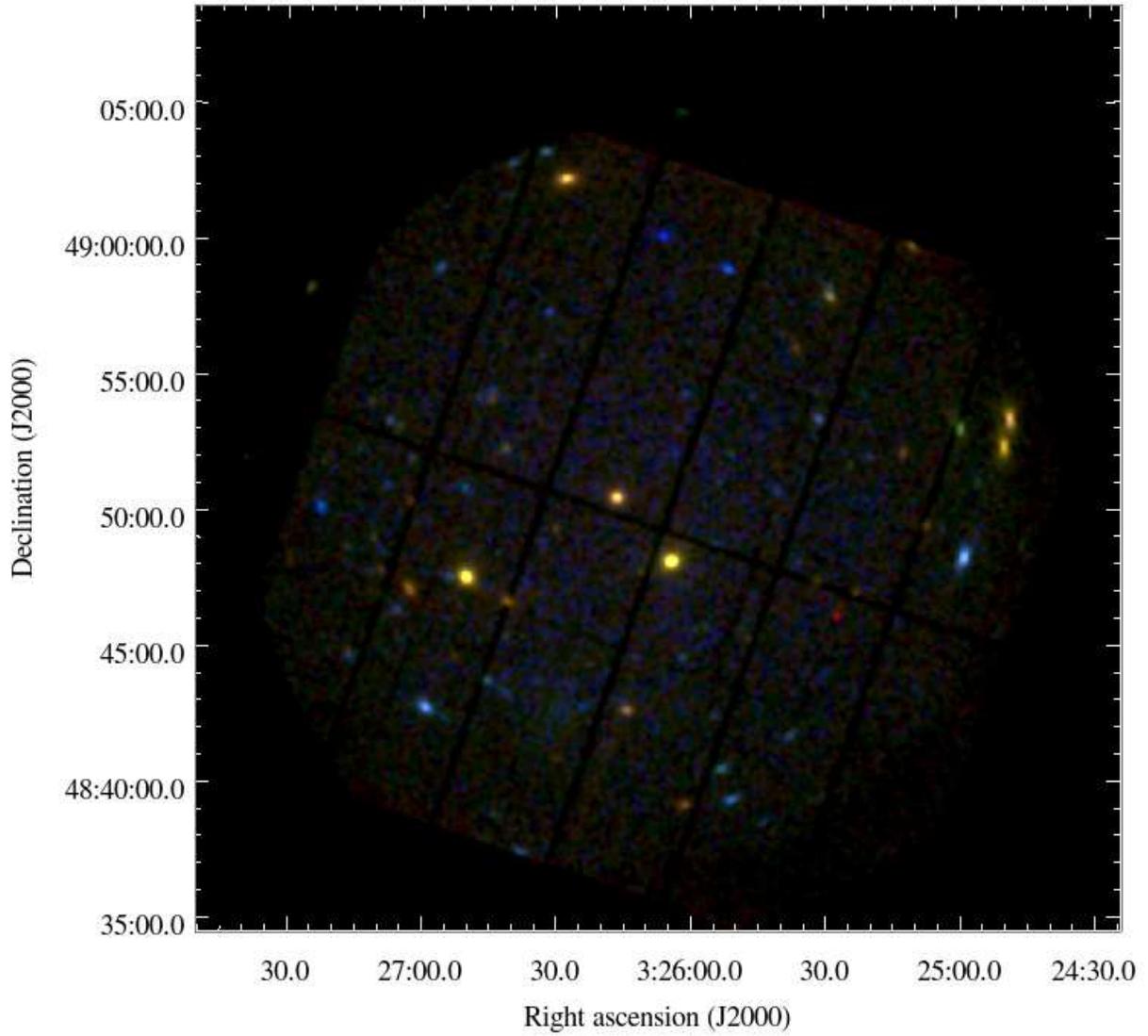}
\caption{Composite PN Mos 1 Mos 2 \xmm\ image within the $\alpha$ Per field. 
The bands used are:  Red$=0.3-1.0$ keV, Green$=1.0-2.5$keV, 
Blue $=2.5-8.0$ keV.   \label{fig1}}
\end{figure}

We carried out  Data Analysis similar to that in Pillitteri, et al. (2004).
We used the standard tasks of  SAS v10.0 to reduce 
the $Observation\/\/ Data\/\/ Files$ (ODF) 
and obtain event tables calibrated in arrival time, energy 
and astrometry. First the events were filtered to be 
within the band 0.3--8.0 keV,
appropriate for the coronal emission we want to investigate. Good time intervals (GTIs)
were filtered out after inspecting the light curve of 
events at energies higher than 10 keV 
\footnote[2]{See http://xmm.esac.esa.int/sas/current/documentation/threads/EPIC\_filterbackground.shtml}
and removing high background intervals 
(with rate thresholds of $>$2.5 ct s$^{-1}$ for PN and $>$0.5 ct
s$^{-1}$ for MOS).  This optimizes the event lists  for the detection
of faint X-ray sources.

%\clearpage

\begin{table}
\caption{List of X-ray sources detected in EPIC-\xmm\ image. The first
  five rows are shown here; the
complete table is available in electronic
  form.}
\begin{tabular}{l l l l l l l} \hline \hline	
ID  & R.A.  	 & Dec. 		 & 	Offaxis 	 & 	Rate  & Err	& Exposure Time  \\ 
	 &  J2000 	 &  J2000		 & 	 \arcmin
&  cts/ks & cts/ks  	&                ks  \\ \hline
1	 & 	03:26:37.9	 & 	48:37:28.9	 & 	15	 & 	15.7	 & 	5.55	 & 	48.3  \\
2	 & 	03:25:43.7	 & 	48:38:32.4	 & 	14	 & 	7.76	 & 	1.63	 & 	50.7  \\
3	 & 	03:26:43.6	 & 	48:38:47.7	 & 	14	 & 	6.75	 & 	1.66	 & 	57.3  \\
4	 & 	03:26:01.3	 & 	48:39:10.1	 & 	12	 & 	21.7	 & 	4.83	 & 	59.9  \\
5	 & 	03:25:50.9	 & 	48:39:22.2	 & 	13	 & 	29.4	 & 	7.41	 & 	57.4  \\ \hline
\end{tabular}
\end{table}

%\clearpage

\begin{table}
\caption{List of sources with 2MASS match and parameters from best fit modeling of X-ray spectra.}
\begin{tabular}{lllllllllll}\\\hline\hline
 ID	&	R.A.	 &	Dec. 	 & 	2MASS ID		&  J	 & 	H   & 	K 	 & 	kT	 &	E.M.	 &	$\log f_X$	 & 	$\log L_X$	\\ 
  	&  J2000	 & J2000	 &		&
 mag	 & 	 mag  & mag	 & 	keV	 & \scriptsize{cm$^{-3}$}	 &
 \scriptsize{ergs s$^{-1}$ cm$^{-2}$}	 & \scriptsize{ergs s$^{-1}$} 		\\ \hline
4	 & 	03:26:01.3	 & 	48:39:09.7	 & 	03260131+4839097	 & 	12.60	 & 	11.92	 & 	11.69	 & 	0.76	 & 	51.7	 & 	-13.8	 & 	28.8  \\
10	 & 	03:26:14.2	 & 	48:42:38.3	 & 	03261419+4842382	 & 	12.48	 & 	11.80	 & 	11.56	 & 	0.89	 & 	51.9	 & 	-13.6	 & 	29  \\
19	 & 	03:26:22.7	 & 	48:44:20.1	 & 	03262270+4844201	 & 	12.77	 & 	12.31	 & 	12.20	 & 	0.86	 & 	50.7	 & 	-14.9	 & 	27.7  \\
22	 & 	03:26:52.6	 & 	48:44:37.9	 & 	03265263+4844378	 & 	11.10	 & 	10.77	 & 	10.76	 & 	0.88	 & 	51.1	 & 	-14.4	 & 	28.1  \\
42	 & 	03:24:58.8	 & 	48:48:14.7	 & 	03245884+4848147	 & 	16.47	 & 	15.69	 & 	14.52	 & 	--	 & 	--	 & 	--	 & 	--  \\ \hline
\end{tabular}
\end{table}

\subsection{Source Detection}  
Source detection was performed using  the algorithm based on wavelet convolution
described in Damiani et al. (1997a,b) and optimized for \xmm \/ 
(Pillitteri et al. 2004).  The threshold used for positive detection
of sources was 4.8 $\sigma$ above local background
fluctuations. 

The full list of source parameters is provided in Table 1 in the electronic version. 
The first few entries are provided in the hardcopy illustrate the content.  The columns
of the table list the source number, right ascension,  declination,  distance off
axis,  source count rate, and effective  exposure time 
(for the summed MOS and PN detectors, assuming a PN--MOS conversion factor of 3.1).   
We detected 102 sources in 0.3-8.0 keV band.

\subsection{Optical Catalogs} 
We have cross-correlated the positions of X-ray sources with the optical catalogs of Prosser (1992),
Randich, et al (1996) and Deacon and Hambly (2004) with a match 
radius of 5$\arcsec$. Only one object of the catalog by
Prosser is in the main field of view of XMM but was not detected. 
Another object from that catalog is on the edge of the XMM field
and was also undetected.  
 Two objects of the Deacon and Hambly catalog
are detected in X-rays, and four stars from the Randich et al. 
catalog are matched with X-ray sources.
 We have used the 2MASS catalog (Cutri, et al. 2003) to find near IR counterparts to the X-ray sources,
finding  39 matches (Fig. 2).
X-ray sources detected on the image (Fig. 1) are a mixture of active young stars from 
the $\alpha$ Per cluster and background objects.  Any star more
massive than M5 would have a K magnitude $\leq$ 13.6 at the age and distance
of $\alpha$ Per (Siess, 2001).  Therefore,   
we assume that sources  without counterparts in the 2MASS catalog are
distant active galactic nuclei (AGNs) which are too faint in the IR.
  Table 2 provides IR photometry for sources identified as stars within 5" of the position  
  of X-ray sources, with
% We have found 
% xxx did we ever cross correlate with Prosser members?  
% or just use 2MASS sources?? 
% We have cross-correlated the detected source 
% list with the 2MASS catalog to identify stellar sources, since background AGN would be much 
% fainter than stars in the near infrared (J, H, and K) in the catalog.  
with source number, coordinates, 2MASS ID, and J, H, 
and K magnitudes in successive columns.   
Parameters derived from fitting (kT, emission measure, and luminosity)
are discussed below in Section 3.2.  Note that in Table 2, X-ray luminosities have been
computed from X-ray fluxes assuming the sources are at the distance of $\alpha$ Per. This is
probably not true for the ``bad L$_X$'' sequence (discussed in Sections
3.1 and 4 below).

\begin{figure}
 \includegraphics[width=\columnwidth]{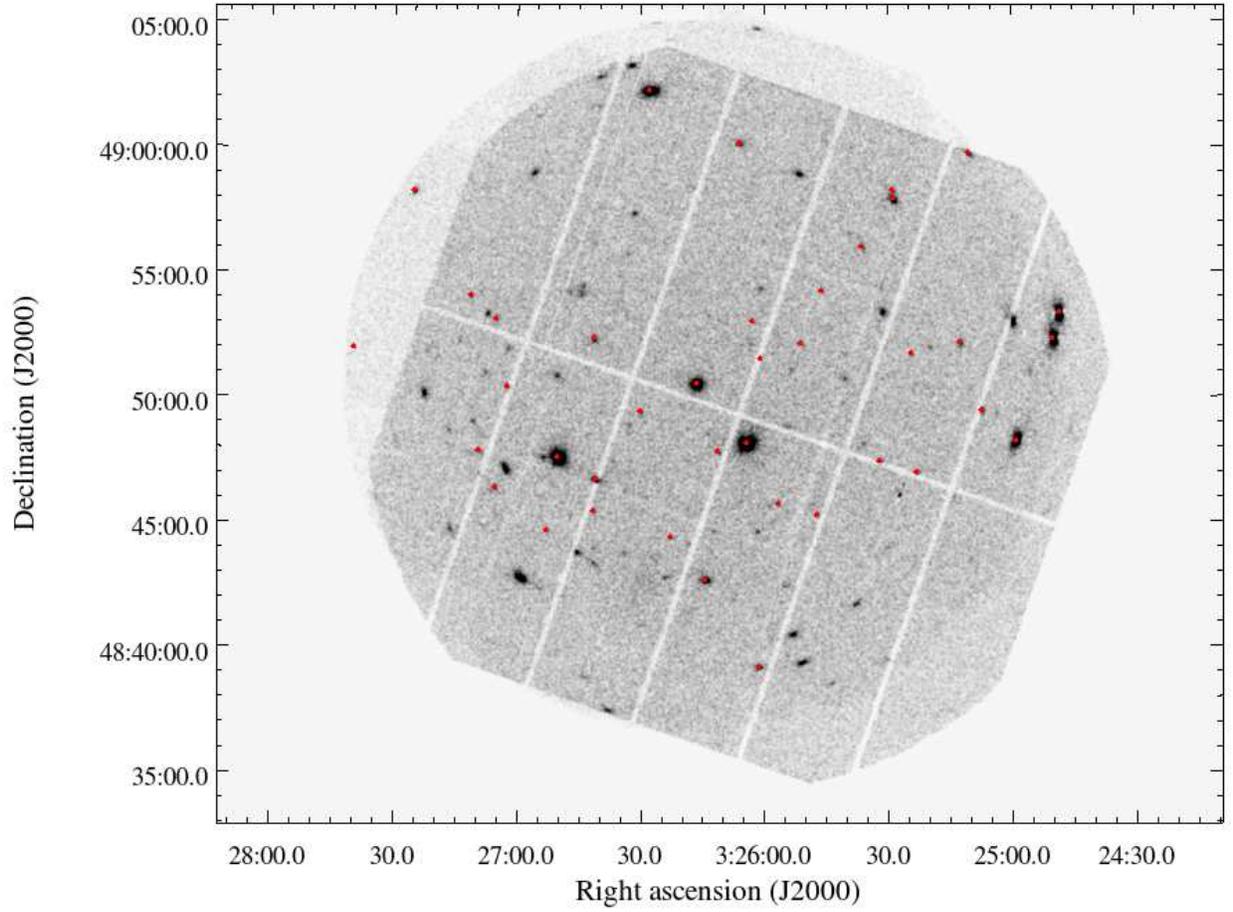}
\caption{The combined PN, MOS1 and MOS2 XMM image with the 2MASS sources indicated in red;  
the  stars from the Prosser catalog are in green.   \label{fig2}}
\end{figure}

\section{Results}

\subsection{The X-ray Luminosity}

The color--magnitude diagram from the 2MASS photometry is shown in Fig. 3.  In it, the 
sources are numbered with the X-ray source ID (Table 1).  The size of the circle is 
proportional to the X-ray luminosity, which   is derived from 
the count rate using a conversion factor from {\it PIMMS} derived from
a  1-T APEC spectrum with kT=1 keV and NH = 10$^{20}$ cm$^{-3}$.  
Overlaid on the figure is the isochrone from Siess et al. (2000)  
for an age of 50 Myr and solar metallicity. 
The figure shows the well-known decrease in X-ray luminosity with
bolometric luminosity along the main sequence.
The Siess tracks allow us to infer masses (see Sect. 3.3).

\begin{figure}
\includegraphics[width=\columnwidth]{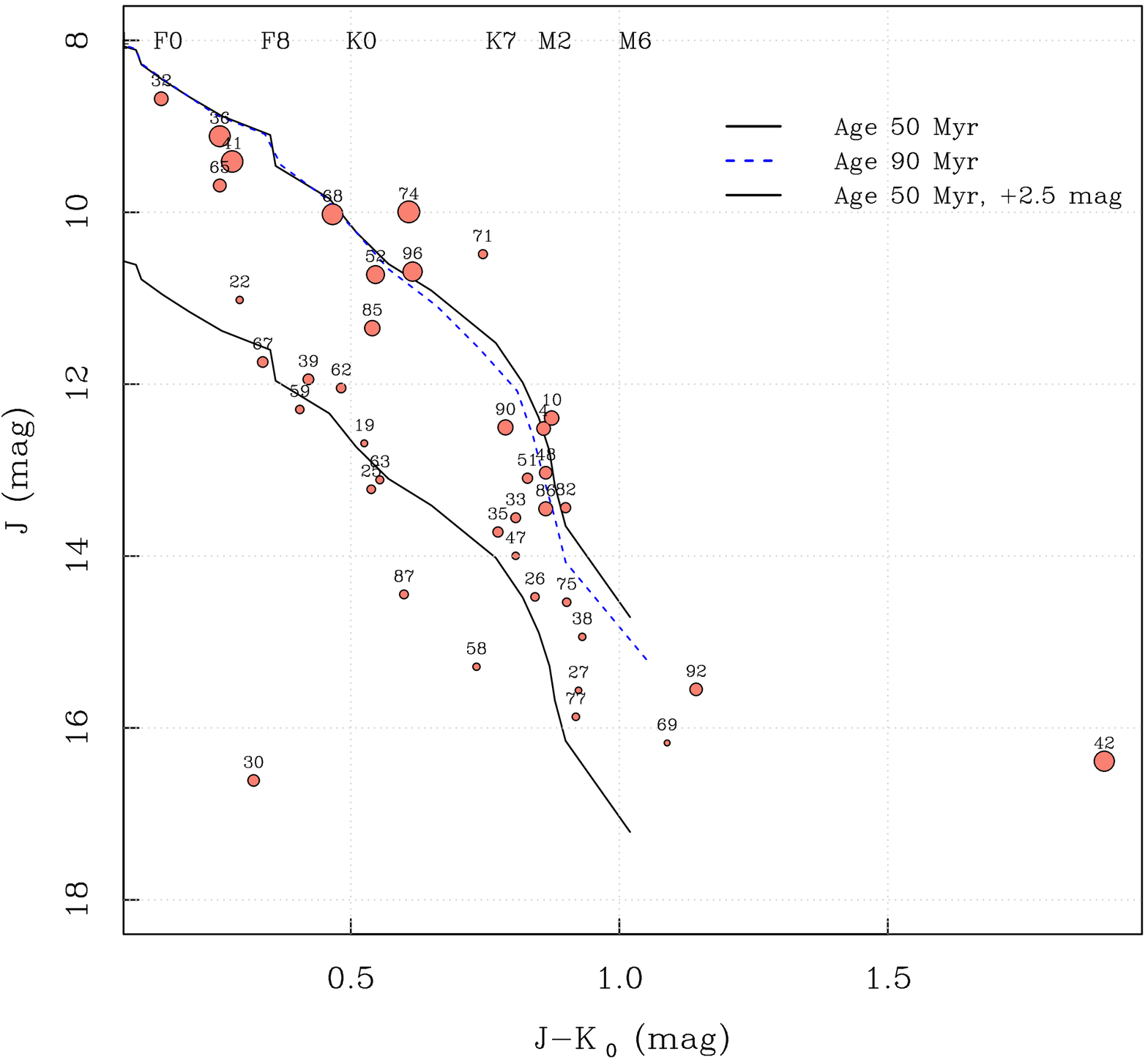}
\caption{The 2MASS J--(J-K)$_0$ color magnitude diagram 
for the stellar X-ray sources.  The circles are proportional to the strength 
of the X-ray signal.  Spectral types for the (J-K)$_0$ colors are indicated 
on the top axis.  The top upper solid  line is the isochrone for 50
Myr age from 
Siess et al. (2000) for the distance of $\alpha$ Per. The dashed line 
 just below that
is the isochrone for 90 Myr.  The solid line at the bottom is the 
same isochrone shifted by 2.5 magnitudes to match the ``bad L$_X$ sequence.
   \label{fig3}}
\end{figure}

A second sequence below the main cluster sequence is present, near the
shifted isochrone (dotted line) in Fig. 3.  This sequence is made up
of sources 19, 22, 25, 39, 58, 59, 62,  63,  67,  and 87 in Table 2. 
For stars fainter than J $\simeq$14 mag, the distinction between the
two sequences is more ambiguous.  However, referring to the morphology
of the color-magnitude diagram in Lodieu, et al. (2012; Fig. 6) the
main sequence is nearly vertical, and we accept the fainter stars as
predominantly belonging to the cluster itself.  
The lower sequence   appears to 
correspond to  sources labeled ``bad L$_X$'',   in PRS Fig. 5.
These sources have a lower X-ray luminosity than the sequence of cluster members.  We draw 
attention to the low-lying sequence in the near IR data (Fig. 3), and will discuss 
the characteristics of these stars in the succeeding sections.   It is
possible, of course, that some of the stars near the isochrone of the
cluster are not in fact cluster members.  Star 71, could for instance
be a foreground star.

We have divided the stars in Fig 3 in a straightforward way to examine the dependencies of 
L$_X$.  There is a significant gap between the cluster main sequence, and the lower sequence, 
the stars  classified ``bad L$_X$''.  Within  the cluster sequence,
stars have been grouped into ``F-G'', ``K-early M'', and ``late M'' according to gaps in the
magnitude distribution.  
Fig. 4 shows the count rate as a function of J magnitude.  The F-G stars (asterisks) and the K-M 
stars (triangles) show the well known progression to lower count rates for cooler stars.
The exceptions are a few of the hottest F-G stars, suggesting that their convective envelopes and 
X-ray production are just becoming established.   The bad L$_X$ stars occur in a separate location, 
with lower count rates and fainter magnitudes, except that the faintest stars in both sequences 
which are mixed in the figure.  Note that this  is
despite the fact that Fig. 3 shows 
that the ``bad L$_X$'' stars are in general 
bluer than the K-M stars in the $\alpha$ Per sequence. 
Similarly, Fig. 5 shows the log count rate as a function of color. The ``bad L$_X$'' stars   
clearly have lower count rates than cluster main sequence stars of comparable colors.
%thus we 
%identify them with the ``bad L$_X$ stars'' discussed by Prosser et al. (1998)   
%(and will refer to them as such below.)  
 Thus low count rate ``bad L$_X$ stars''  have  properties distinct from cluster members
in both near IR photometry (Fig. 3) and in X-rays.

\begin{figure}
 \includegraphics[width=0.6\columnwidth,angle=-90]{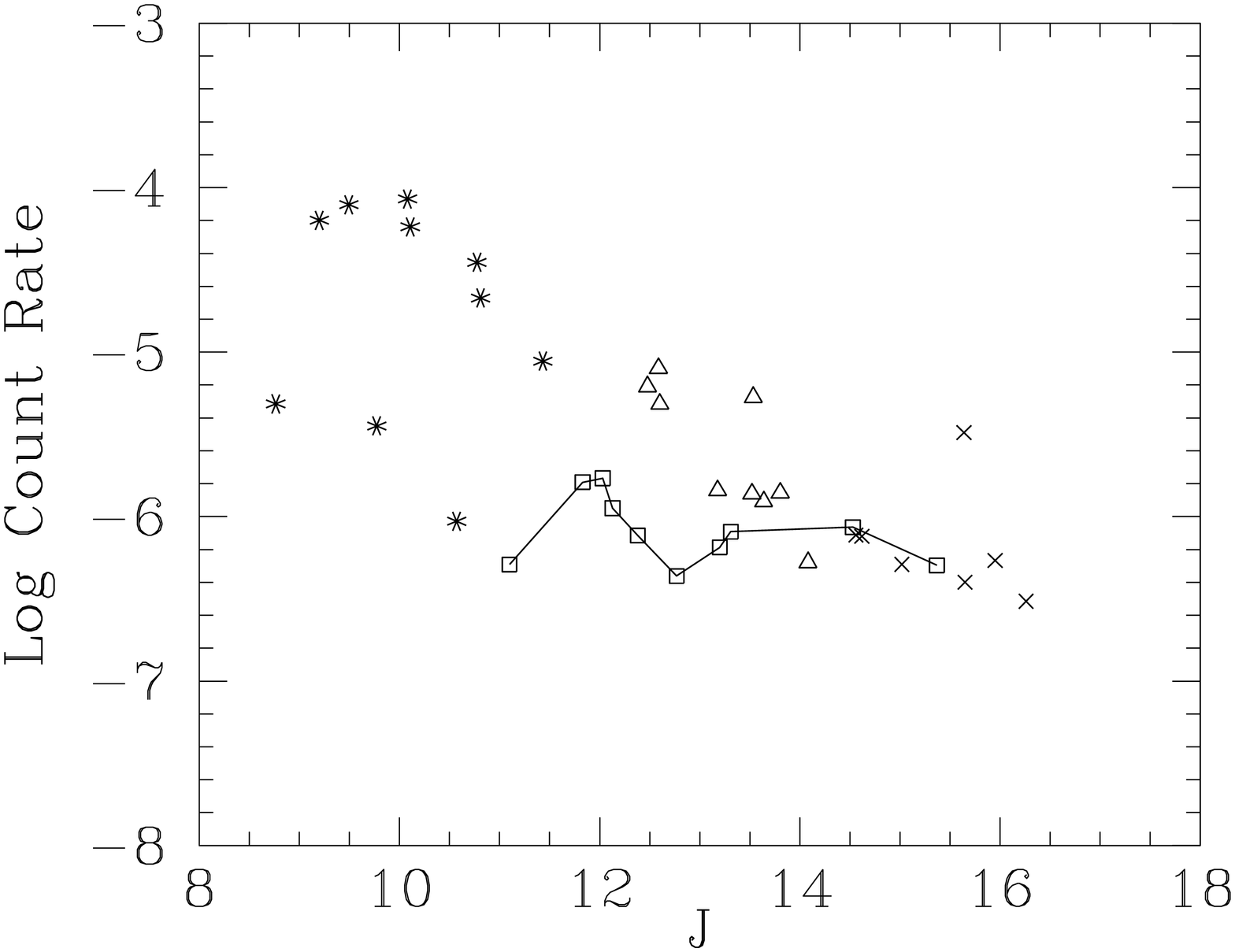}
\caption{The log of the count rate as a function of J magnitude.  Symbols are
as follows: F-G stars: *; K-early M stars: triangles; late M stars: x; ``bad L$_X$'' 
stars: squares.  To emphasize the location of the ``bad L$_X$'' stars, they have been
connected in order of magnitude.  Two distinct sequences are seen,
merging at the faintest stars.  J is in magnitudes, count rate is
counts per ksec.
   \label{fig 4}}
\end{figure}

\begin{figure}
 \includegraphics[width=0.6\columnwidth,angle=-90]{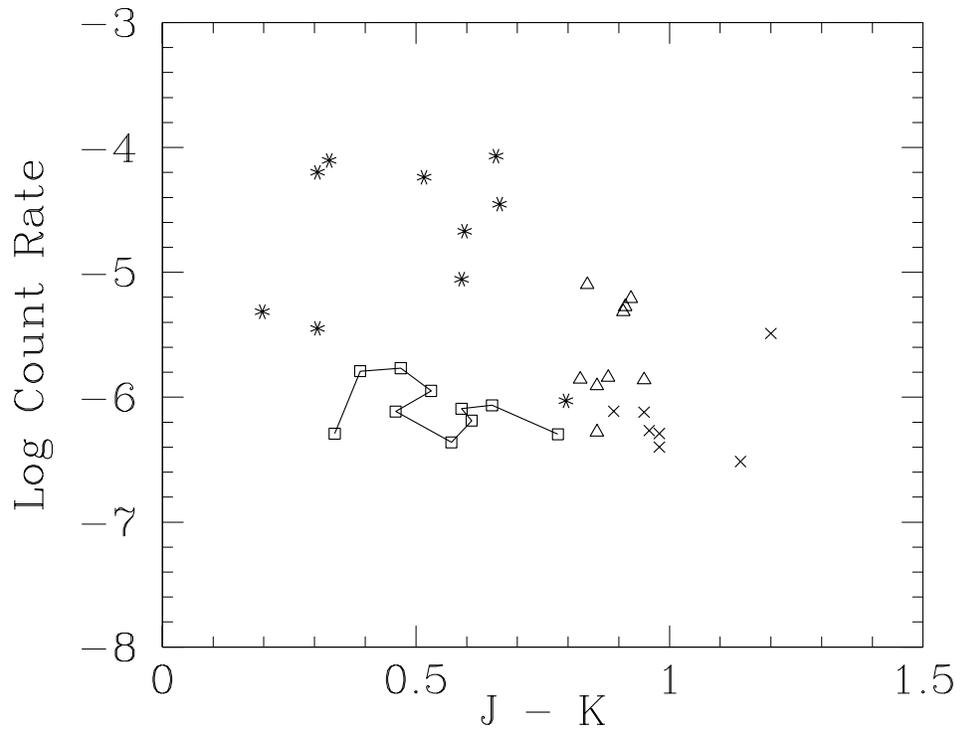}
\caption{ The log of the count rate as a function of J-K color.
 Symbols are the same as Fig. 4.  The ``bad L$_X$'' stars clearly have 
lower count rates than cluster main sequence stars of the same color.  
  J-K is in magnitudes, count rate is
counts per ksec.
  \label{fig 5}}
\end{figure}

\begin{figure}
\includegraphics[width=0.6\columnwidth,angle=-90]{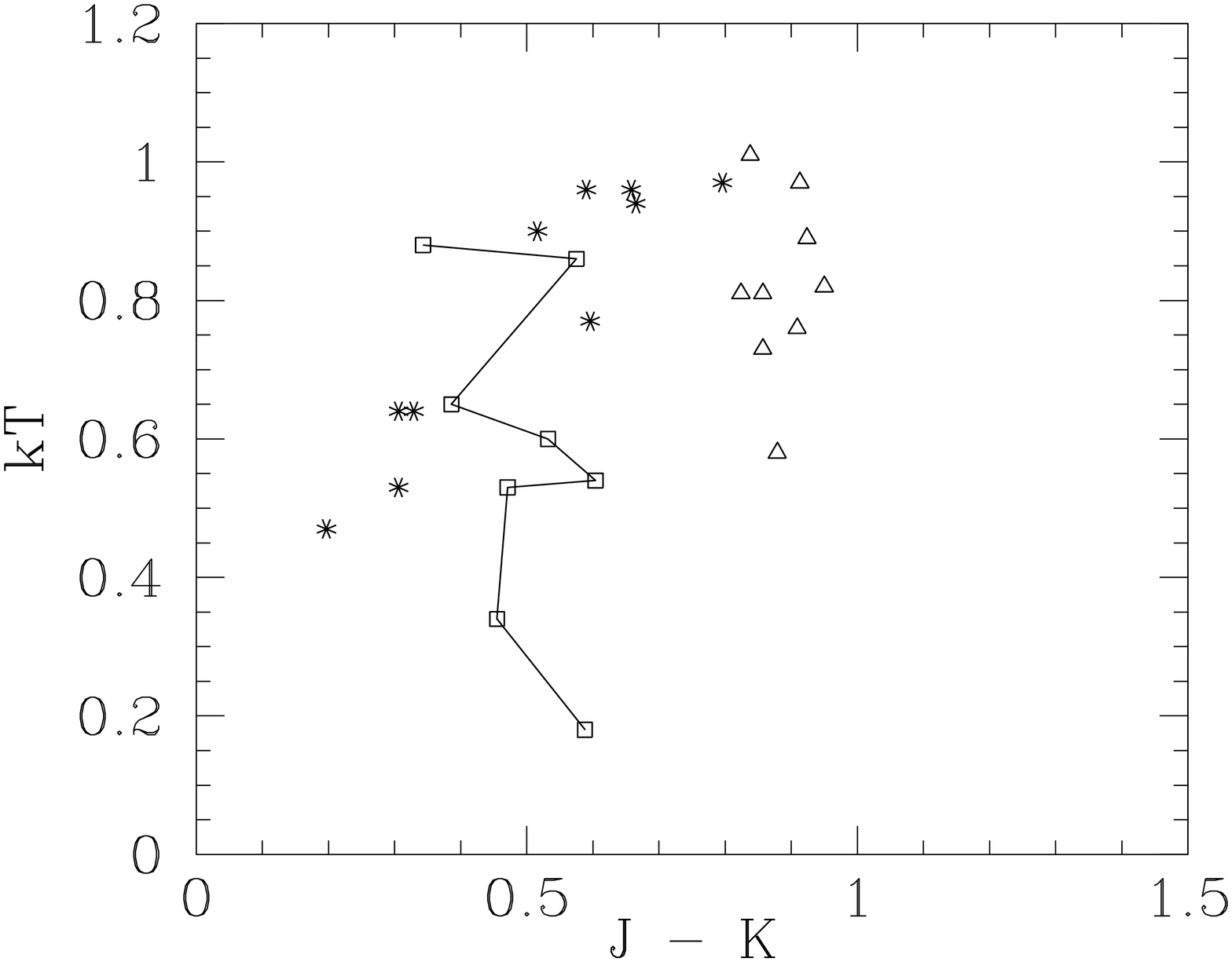}
\caption{ The temperature kT (keV) as a function of the J-K color.  Symbols are the same 
as in Fig. 4, except that the ``bad L$_X$'' stars are linked are
linked in order of kT.  J-K is in magnitudes, kT is keV.
   \label{fig 6}}
\end{figure}

\begin{figure}
\includegraphics[width=0.6\columnwidth,angle=-90]{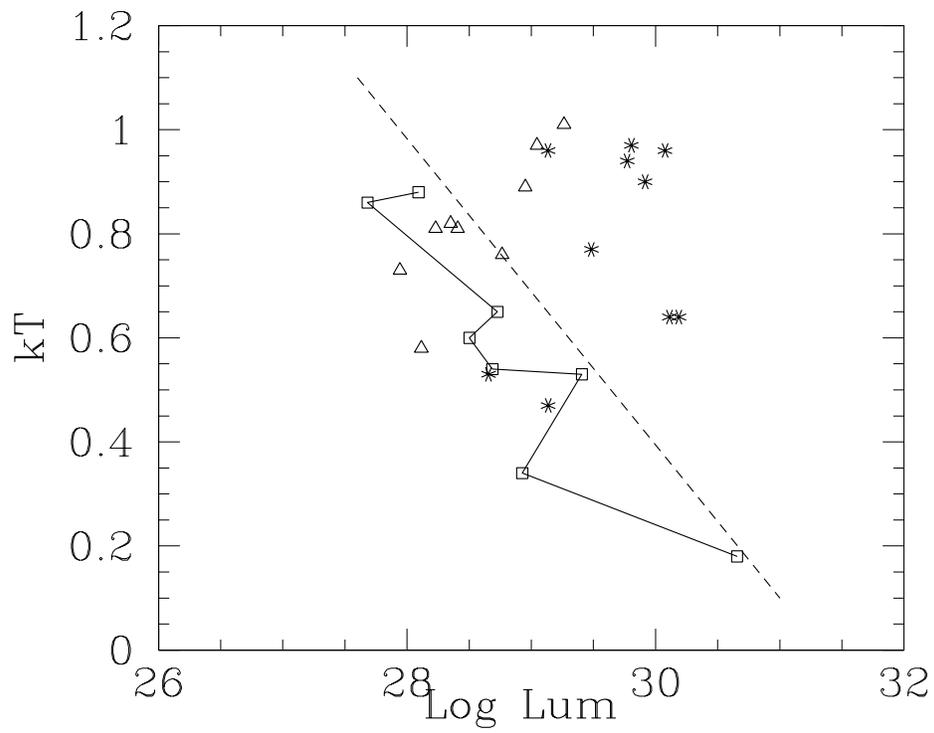}
\caption{ The temperature kT (keV) as a function of X-ray luminosity.  Symbols are the same 
as in Fig. 6.   The dotted line is the upper envelope below which 
the bad L$_X$ stars fall.  Luminosity is in ergs cm$^{-2}$ s$^{-1}$.
   \label{fig 7}}
\end{figure}

\subsection{Spectra}

One of the most important parameters to examine for the cluster stars is the temperature derived
from the X-ray spectra.  In order to use temperature from a large number of sources, we have made 
 1 temperature fits of {\it APEC} models to the spectra with
 {\it XSPEC} software (ver 12.6). 
These temperatures are listed in Table 2.  For low countrates, we did
not fit spectra, and the temperature, emission measure, flux and
luminosity columns in Table 2 are blank. These sources are omitted
  from figures.  (Luminosities in Fig. 3 were
derived as in section 3.1 assuming kT = 1 keV.)  
Fig. 6 shows the temperatures for the groups of stars as a function of the J-K color.  The late M 
stars in Figs 4 and 5 are all too weak to have a reliable temperature determination.  
All groups (the F-G stars, the K-M stars and the bad L$_X$ stars) have a range to temperatures.  
However the mean X-ray temperature kT for the  
bad L$_X$ stars (0.57 keV) is lower than the mean for the other 2
groups (0.80 keV).  
Fig. 7 shows the X-ray temperature as a function of luminosity for all groups.  The luminosity is computed 
on the assumption that all the stars are at the cluster distance.  Both the F-G and K-M stars show 
a range of temperatures along 
with the well-known decrease in X-ray luminosity for the lower mass
stars. These two groups exhibit minimal overlap in the figure and a
large dispersion.   The dotted line indicates the upper envelope of the
bad  L$_X$ stars, which do not in general reach the highest luminosities of the other groups.

To illustrate the effects of different source temperatures, Fig. 8 shows the spectra of two
strong sources, Src 68 and Src 36.  Src. 36 is an F star (J = 9.2 mag) with a single 
temperature fit
(Table 2) kT = 0.64.  Src 68 has J = 10.11 mag, and kT = 0.9.  The softer spectrum
 of Src 36 is 
clearly apparent in Fig. 8.

\begin{figure}
 \includegraphics[width=0.6\columnwidth,angle=-90]{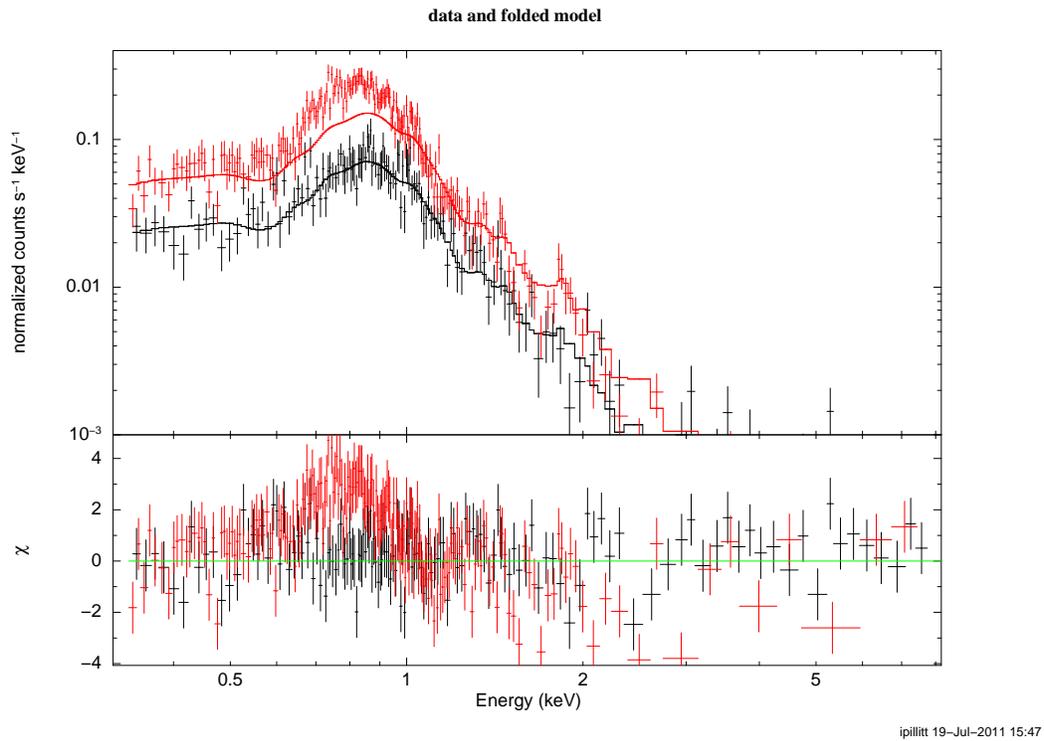}
\caption[]{Spectral fits of sources 68 and 36. Top:  Src 68 is in black; Src 36 is in red.  
The red line is the spectral fit to Src 68 shifted upward for comparison with Src 36.  Excess
soft flux in the spectrum of Src 36 is clearly seen in comparison with the fit to Src 68. 
Bottom: Residuals from the fit in the same colors as the top.  
   \label{fig 8}}
\end{figure}

%In addition we have produced 2 temperature fits for the strongest
%sources. 
For about half the strong sources, the $\chi^2$ was
satisfactory for a single temperature fit.  For the other sources, 
Table 3 lists the temperatures for 2 temperature fits, together with
the off-axis distance and source counts. Source counts  include all 3
cameras (MOS1, MOS2, AND PN) taking into account the different
effective areas of the 3 cameras. 
%xxx Ignazio:  source counts is the exposure time times the count rate
%in the source table (table 1):  is that right?? xxx 
Fig. 9 shows the results.  A correlation between the
two temperatures is present, as found by  Wolk, et al. (2005) and Briggs and Pye
(2003).  In order to investigate this relation as a function of age,
%Furthermore the relation between the two components changes
%with the age of the system (e.g. G\"udel, Guinan, and Skinner 1997;
%Favata and Micela 2003).
Fig. 9 includes data from  the
1-3 Myr Orion Nebula Cluster (ONC) COUP project
(Wolk, et al. 2005),  the 100 Myr  Pleiades (Briggs and Pye 2003) 
and 30 Myr NGC 2547
(Jeffries, et al. 2006) as well as fits to XMM ONC sources (Table 4) to explore
this with XMM and Chandra data.  ONC Chandra data were taken from Wolk, et
al. for the ``characteristic flux'' (their Table 4), omitting the data
for the 4 stars flagged as poor fits.  Their fit to that data 
(kT$_2$ = 2.14 $\times$ kT$_1$ + 0.660 keV) is also
included in Fig. 9.   Data for the Pleiades were taken from the XMM
observations of Briggs and Pye (2003), using the two temperatures from
the PN (their Table 3), since only one source had determinations for
several instruments, and they are all similar. 
\footnote[3]{Daniel, Linsky, and
Gagne (2002) also fit Chandra spectra of several sources, but kT$_2$
was fixed in their solutions.}   We also include 9 sources from the 30 Myr
cluster NGC 2547 (Jeffries, et al. 2006).
 Data for NGC 2547 is taken from the 2
temperature fits in Jeffries, et al. (2006; their Table 5), omitting  sources
where kT$_2$ is described as
unconstrained in the fits.   Sources which are late B
stars (\#7 and \#8) were also omitted. However, their two 
temperatures match those of the cool stars very well,
confirming that the X-rays from these sources come from a low-mass
companion.

%\clearpage

\begin{table}
\caption{Two Temperature Fits for the $\alpha$ Per Cluster }
\begin{tabular}{lllrr}\\\hline\hline 

Src 	&	kT$_1$	 &	 kT$_2$	& Off Ax & Cts \\
 
 & 	keV &	 keV	& $\prime\prime$ &   \\

4   &	0.18	  &  0.88	  &  12 &   289  \\ 	
36  & 	0.5	  &  0.86	 &  8.1 &  7030	 \\
41  & 0.46	  &  0.9	 &  3.5 & 10200	 \\
52  & 0.24	  &  0.98	 &  1.8  &  3170	 \\
68  & 	0.77	 &  1.42	 & 13  &  2100	 \\
74  & 0.81     &  1.48	 &  13 &  2380	 \\
85  & 0.74	 &  1.52	 & 9  &   651 \\	
90  & 	0.81	 &   2.6	 &  12 &  279 \\	
96  & 0.57	 &  1.24	 &  11 & 2520 \\	

\hline

\end{tabular}
\end{table}

%\clearpage

\begin{table}
\caption{Two Temperature Fits for the ONC (XMM) }
\begin{tabular}{lllr}\\\hline\hline 

Obsid & Src 	&	kT$_1$	 &	 kT$_2$	 \\
 
 & &	keV &	 keV  \\

 0212480301 &	304  & 0.97  &	2.49  \\
 0093000101  &   281 &  0.78  &    1.86  \\
 0212480301  &   221 &  0.30  &    1.58  \\
 0212480301  &   237  & 0.51  &    1.33  \\
 0212480301  &   280 &  0.81  &    1.57  \\
 0093000101 &    281 &  0.78  &    1.86  \\
 0093000101  &   229 &  0.84  &    1.76  \\
 0093000101  &   200 &  0.73  &    1.86  \\
 0093000101  &   194 &  0.96  &    2.14  \\
 0093000101  &   196 &  0.25  &    1.30  \\
 0093000101  &   309 &  0.20  &    0.99  \\
 0093000101  &  132 &  1.04   &   3.71  \\
 0093000101  &   158 &  0.25   &   1.02  \\

\hline

\end{tabular}
\end{table}

\begin{figure}
\includegraphics[width=0.6\columnwidth]{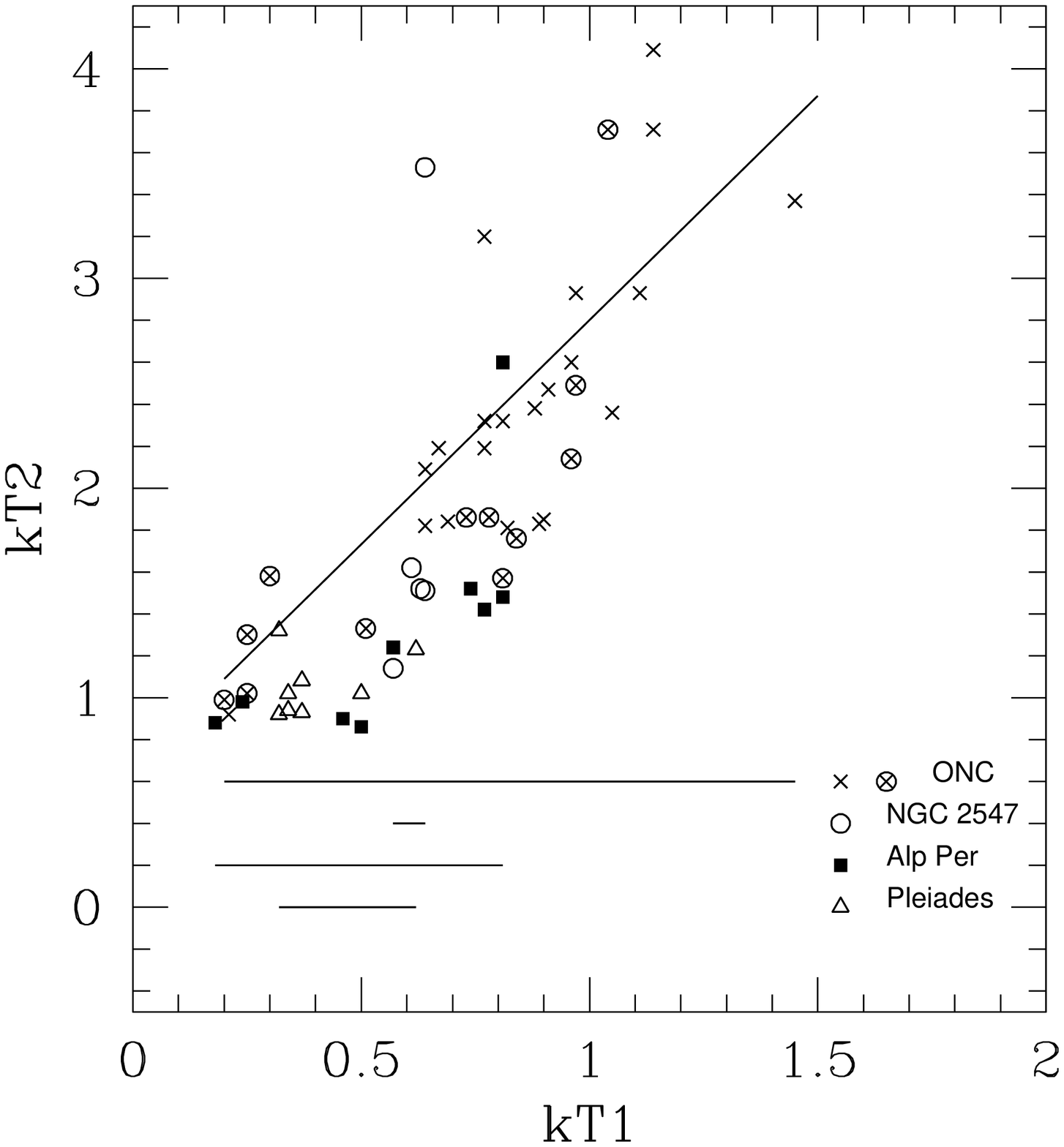}
\caption{ The two temperatures from spectral fits, kT$_1$ and kT$_1$.
$\alpha$ Per stars are shown by filled squares, Orion Nebula Cluster
  sources from Wolk et al. (2005) by x's,  XMM  Orion Nebula Cluster sources
  (Table 4) by circled x's;
Pleiades stars by open triangles and NGC 2547 by
  open circles.  The line is  the fit to the Wolk et al. Orion data.  Beneath the
plot the range of the kT$_1$ for each cluster is shown, from the youngest
cluster (the ONC) to the oldest (the Pleiades).  kT$_1$ and kT$_2$ are
in keV.
   \label{fig 9}}
\end{figure}

Fig. 9 shows that both the kT$_1$ and kT$_2$ temperatures are
smaller on average for the 50 Myr  $\alpha$ Per stars than 
for the younger  ONC stars (1-3 Myr; Megeath, et al. 2012). 
This is expected as a result of the decrease in stellar
activity as stars age.  (The single point for $\alpha$ Per with kT$_2$
$>$ 2 is for a source which is both weak compared to the others  
and off axis, and hence has a significant systematic error in addition to
the measurement error.)  The Pleiades stars (100 Myr)
fall among the $\alpha$ Per stars, but the hottest of both 
kT$_1$ and kT$_2$ for  Pleiades stars 
is cooler than those for the $\alpha$ Per stars.  
Similarly, the NGC 2547 stars (30 Myr) largely overlap the $\alpha$
Per stars.  
%(The star with the large kT$_2$ has a very large
%uncertainty on that temperature, and will be omitted in further
%discussion.)
  The lines at the bottom of Fig. 9 indicate the range
of kT$_1$ values ordered from the youngest cluster (ONC) to the oldest
(the Pleiades).  
This shows the decrease in stellar
 activity as stars age. Once on the Zero Age Main Sequence
(ZAMS), rotation and hence dynamo 
activity decreases as stars spin down.  Jeffries et al.  discuss activity at 30 Myr in
 NGC 2547 for G, K, and M stars.  At this age, solar mass stars have
 just contracted to the ZAMS.  They argue that coronal temperatures
 decrease up to this point as gravity increases.  The decrease in the
 harder component in solar mass stars was previously noted  
by G\"udel, Guinan, and
 Skinner (1997) and subsequently by Telleschi, et al. (2005)

There is  a correlation between kT$_1$ and kT$_2$ in Fig. 9, particularly
for  $\alpha$ Per and the ONC.  There is a suggestion, however, that
the relation for $\alpha$ Per is less steep than for the ONC.  That
is, even for the same soft component temperature kT$_1$, the hard
component decreases as stars age.  The Pleiades stars (100 Myr)
 cluster among the $\alpha$ Per stars, although the
coronae appear generally cooler than the $\alpha$ Per sample and there
appears to be much less differentiation among the individual stars, with 5 of
the 8 being well described as kT$_1$ = 0.35 $\pm$ 0.050 keV and
kT$_2$ = 0.950 $\pm$ 0.100 keV.
    Due to this clustering no slope can be determined for the Pleiades
stars. In fact the concept may no longer be relevant at that age if
  the hot component is gone. Typically
 older cool stars have a corona that is dominated by a
one-temperature cool plasma.
If the cool component at 100 Myr has already reached this asymptotic minimum,
 then these stars may be  best described merely a single temperature
which is a function of time.

%COMMENT:
%Jeffries  2006 argue *T *=E2=88=9D *Bg\^*=E2=88=921/2.     on the radiativ=
%e track ((alpha
%Per and Pleiades this simplies further to
%*T *=E2=88=9D *B*.   I THINK the T in this case is the hot component , but =
%I need
%to think about that more.    if we know the B goes down with time
%(Ignazaio? Help!) then this is solved.

%There is a suggestion, however, that
%the relation for $\alpha$ Per is less steep than for the ONC, which is
%consistent with the data for the Pleiades and NGC 2547.  That
%is, even for the same soft component temperature kT$_1$, the hard
%component decreases as stars age.   
This discussion, of course, is produced from
a sample for all clusters that is biased in the sense that the two
temperature fits are only possible for  the strongest X-ray sources.  
In addition, in the $\alpha$ Per cluster, a number of strong sources
are satisfactorily fit with one temperature.  Furthermore, the samples
from different clusters may contain a different distribution of
spectral types, again creating possible bias. 

\subsection{X-ray Luminosity Distribution (XLD)}

Although the XMM field covers only a small part of the $\alpha$ Per cluster, 
the exposure is deeper than the ROSAT images.   We 
derive an X-ray luminosity distribution (XLD) to compare with the ROSAT results 
for the full cluster area using the {\it ASURV} software package. 
  As discussed in Sect. 
2.2, there are two sequences.   The brighter sequence is 
from the cluster itself.  There is 
also a fainter sequence, corresponding to the ``bad L$_X$'' stars, which will be 
discussed further in Section 4.   
%Fig. 3 guides the division of the
%cluster sequence into categories to explore the XLD.
The XLD is derived from the upper cluster sequence. 
Fig. 3 shows that it contains  a natural divide in the J--(J-K)$_0$ diagram, 
with a gap in both color and magnitude.  The sources  32, 36, 41, 52, 65, 68,
71, 74,  85, and 96 in Table 2
belong to the earlier side of the low-mass main sequence stars (G-K stars).  
Others in that sequence are 
assigned to the cooler low-mass star group (M stars).  The division in (J-K)$_0$ comes at the 
end of the K dwarf color, so the first group contains F, G, and K stars. 
Since the number of sources in the part of the cluster we are sampling is small, we are not 
able to subdivide this group further.  The cooler 
group contains M stars.  Since we do not have have a
list of authenticated faint cluster  members, we do not have upper limits for nondetections
of faint members. 
Fig. 10 provides the results.  The luminosity distribution for the F, G, and K stars is similar to
the results from Randich, et al. (1996) for the full cluster.  For instance, the mean log L$_X$
from that study is 29.63 ergs sec$^{-1}$ (F dwarfs), 29.74 ergs sec$^{-1}$ (G dwarfs), and 
29.56 ergs sec$^{-1}$ (K dwarfs), comparable to the blue line in Fig. 10.  For the M dwarfs,  the
deeper XMM data provide a distribution that continues more smoothly to lower luminosities than 
the Randich, et al. distribution which has a mean log L$_X$ if 28.96 ergs sec$^{-1}$.  
For completeness, we derive an XLD for the ``bad L$_X$'' distribution as though it were at the
same distance as the cluster even though this is likely an incorrect assumption (see 
Section 4).  Thus, the XMM results for the $\alpha$ Per cluster
 are in agreement with the previous
ROSAT results for the XLD for F, G, and K stars, but they extend the
XLD to fainter sources for M stars.  Fig. 10, for instance provides a
basis for inferring the properties  of possible low-mass companions of
Cepheids (Evans, et al. 2012), necessary for 
 calculating X-ray exposure times.

\begin{figure}
 \includegraphics[width=\columnwidth]{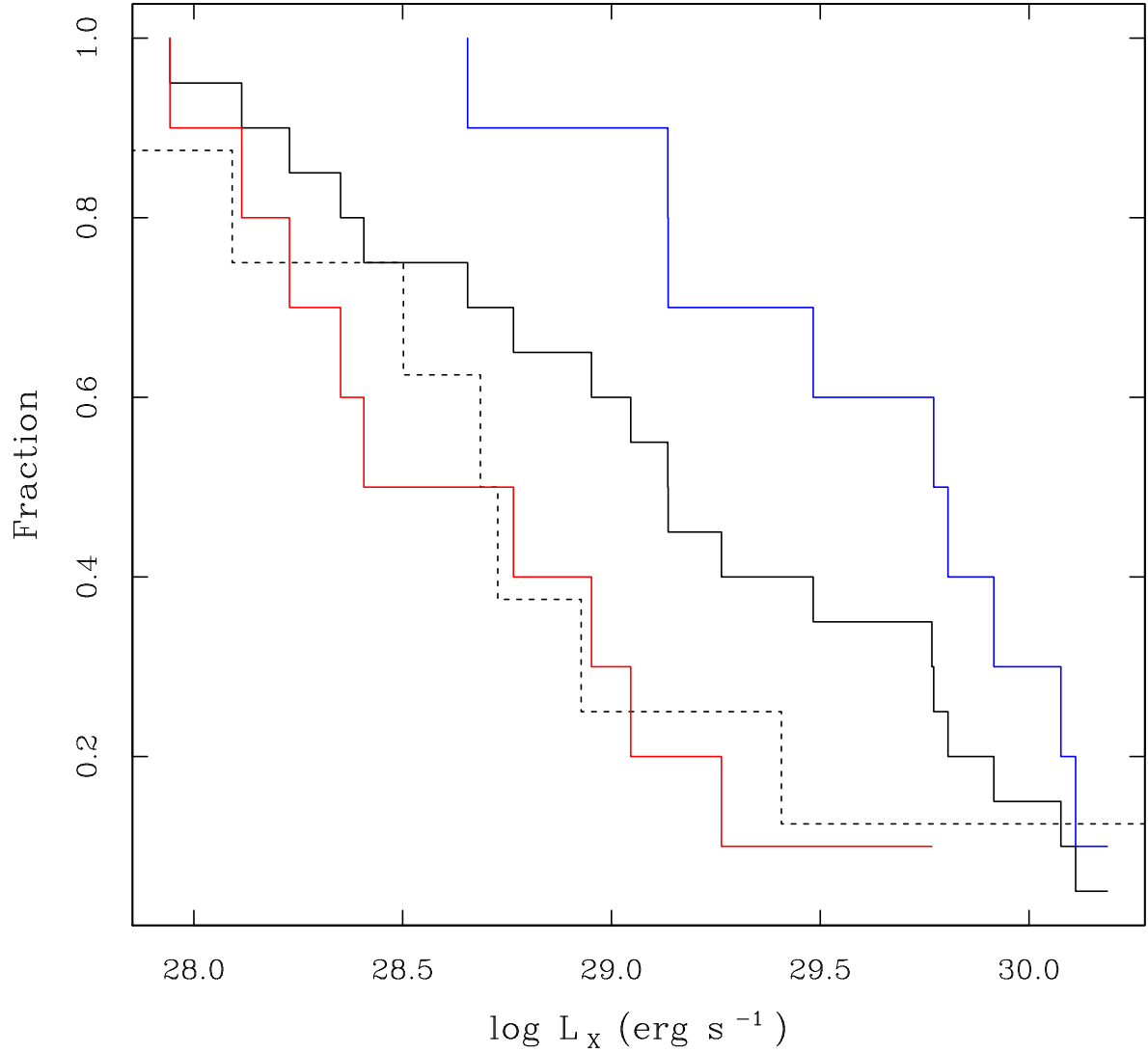}
\caption{The X-ray Luminosity Distribution (XLD).  The black solid line is for the all 
spectral types; the blue solid line is for the hotter of the low-mass main sequence stars; 
the red solid line is for cooler main sequence stars; the dotted line is for the 
``bad L$_X$'' stars, assuming they are at the distance of the cluster (see text for 
discussion).
   \label{fig 10}}
\end{figure}

\subsection{X-ray Time Variability}

We have examined the light curves of the sources to look for time
variability, since  
flares are expected at the young age of $\alpha$~Persei cluster.
However because most of the sources have relatively low counts and
60 ksec is a short interval, we discuss here only
six bright sources that show possible variability in inspection by eye.
 Fig. 11 shows the PN light curves of these bright sources, all 
of which belong to the ``cluster'' sequence.
%In this sequence there are 26 X-ray sources, leading to an 
% estimate of the  rate of variability of $\ge23\%$.  Note that this 
% estimate is biased by faint sources that do not show 
%variability because of the poor count statistics. 
  Source
36 shows a  smooth rise of about  1 $\sigma$ level and a flare-like event
   at 43 ks. The flare has peak rate   $~$50\% 
larger than the pre-flare rate and a
   significance $>$ 2 $\sigma$. The duration is  3-4 ks.
   Other possible small flares  are visible at
   about the 1$\sigma$ level.
% A dip in the rate is present at $\sim12ks$ but its nature 
%is unclear. The dip is much less pronounced in MOS light curves. We cannot exclude it is
%a instrumental feature like a loss of telemetry. Another remote possibility is that this is
%a real feature  suggesting an obscuration from a large body, opaque in X-rays, similar
% to that reported by Briggs et al. 2003 for a star in the Pleiades. 
%   (A dip is visible at 12 ks at a level of ~2sigmas, it is not clear if it is present
%   in MOS ccds so I would not trust too much to this feature, a loss of telemetry could explain this)
Sources 41, 68, and 96 are largely constant, with 
possible variability at the 1 $\sigma$ level.
Source 74  shows a possible sequence of flares, the first  at $\simeq$13ks
with  a duration of 8 ks and peak rate about 80\%  larger than minimum observed rate.
 Source 90 is interesting because it has a very low count-rate in the
 first half of the observation,
   then a slow increase and a steeper fall within 10 ks. The peak
   rate in this event is about 8 times the quiescent rate.
In summary, in the brief window of the observation, several sources
show evidence of low level variability, but only one has a significant
flare.

\begin{figure}
 \includegraphics[width=\columnwidth]{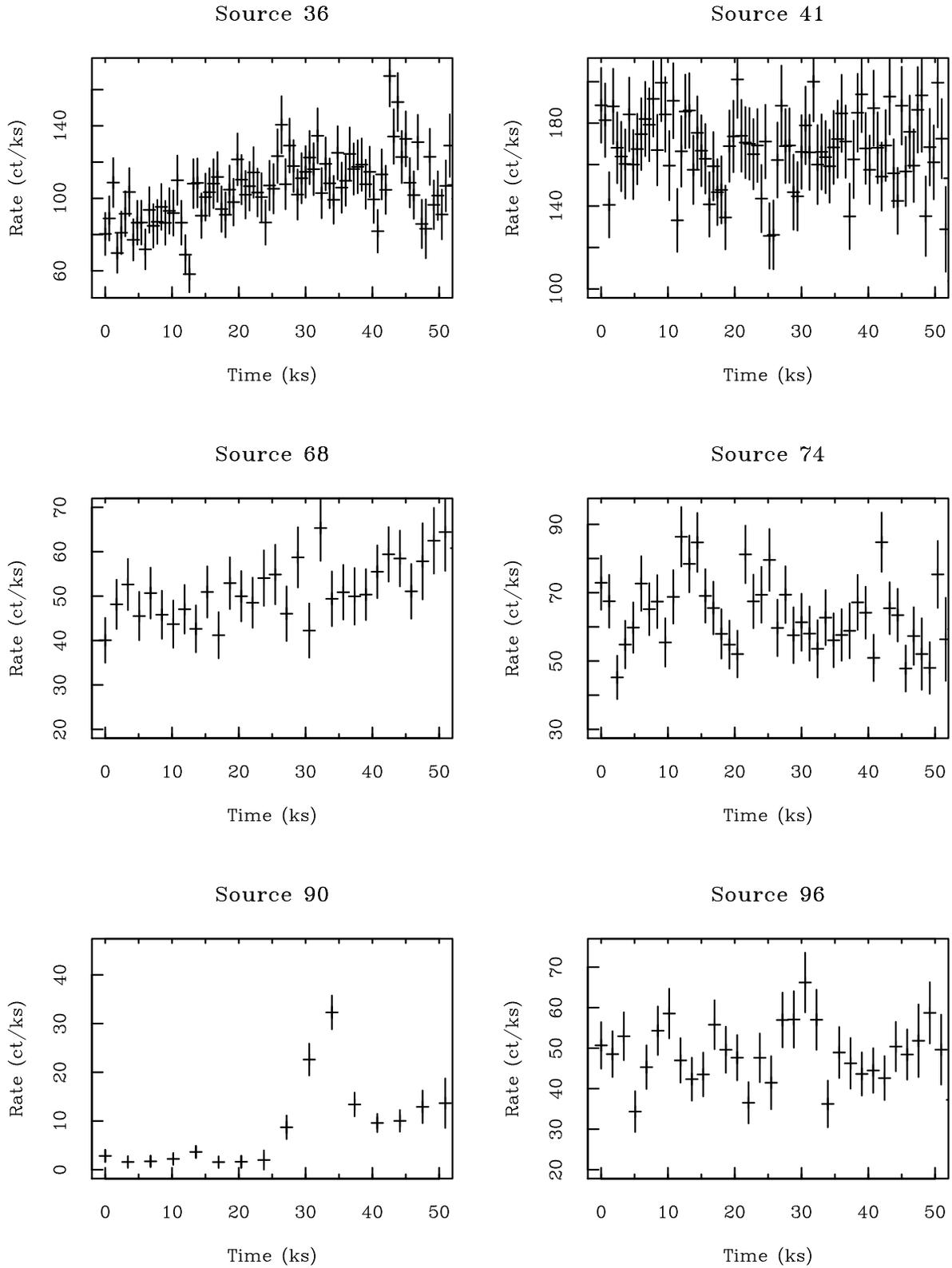}
\caption{Light Curves for a selection of brighter sources. See text
  for discussion.
   \label{fig 11}}
\end{figure}

%\section{ Comparison with ROSAT Results}

%probably already done in the XLD section 

\section{Discussion: Bad L$_X$ Sources}

We have used the deep XMM exposure of the $\alpha$ Per cluster to sample 
the XLD of a section of the cluster.  In particular we have added a 
sample of M stars to the previous work.  This confirms that the distribution 
fits nicely between the youngest clusters (e.g. the Orion Nebula Cluster) and 
the older Pleiades.  

An unusual feature of both the ROSAT and the XMM images of the cluster is the 
additional sequence, dubbed the ``bad L$_X$'' stars by PRS. 
It was first identified in the ROSAT images, however it is clearly evident in the 
near IR color-magnitude data of the XMM X-ray sources (Fig. 3) as well.  The simplest 
explanation is the geometric one, that the ``bad L$_X$'' sequence comes from a 
more distant grouping behind the $\alpha$ Per cluster.  Interpreting Fig. 3 in this
way, the sequence is  2.5 magnitudes fainter than the $\alpha$ Per 
sequence as shown by the shifted isochrone in Fig. 3, 
which translates into a factor of 3   further in the
distance, making it about 500 pc away. This ignores possible  additional reddening at that
distance.  However, the faintest pair of stars in the ``bad L$_X$''
sequence are in the late-K and M region where the main sequence in
Fig. 3 is nearly vertical.  Their (J-K)$_0$ is not consistent with
significant additional reddening as compared with the cluster sequence. The
increase in the  `bad L$_X$'' distance of 3   implies an addition of 1.  to 
the log L$_X$  or log count rate.  Fig. 5 provides a good check on this possibility. 
Even if the log count rate were adjusted by this amount, it would still
fall a little below 
the $\alpha$ Per sequence for the same J-K.  This suggests 
 that the grouping may  also be a little older than the 50 Myr 
$\alpha$ Per cluster.  There is evidence for this in Fig. 6, in that the average
kT (in the relevant J-K band) is a little smaller than for the $\alpha$ Per stars 
with the same colors.  Thus there is a suggestion in Fig. 6 that the
``bad L$_X$'' stars may be slightly older than the $\alpha$ Per
stars.  However there are only 8 stars and several parameters involved
(distance, reddening, age/temperature) so this is not a conclusion,
rather a possible interpretation of the available data.  
For a second comparison, in Fig. 10, if the
 estimated addition to the ``bad L$_X$''  of 1.  is subtracted from 
the log L$_X$ of the F-G-K stars of the cluster (blue line), it
would be moved nearly to the log L$_X$  LXD (dotted line). It is not
surprising from Fig. 3 that faint M stars are not detected at the
distance of the ``bad L$_X$'', and hence the F-G-K stars (blue line)
are the best comparison at that distance. 
While the small number of``bad L$_X$'' stars preclude firm
conclusions, we at least offer a plausible interpretation of a group
about the same age as the $\alpha$ Per cluster or possibly a little
older.

% In sum, the ``bad L$_X$''
%group is probably roughly the same age as the $\alpha$ Per cluster or
%a little older.  

PRS remarked that the ``bad L$_X$'' stars are frequently classified as nonmembers
of the $\alpha$ Per cluster on the basis of radial velocities.  Of the 7 ``bad L$_X$''
in their Table 4 with radial velocities, 5 are listed as nonmembers on the basis
of radial velocities, and for the other two, the membership status is questionable.

 There is one restriction on the distance of the ``bad L$_X$''
stars, which is that they cover a large part of the  $\alpha$ Per area surveyed by PRS.
That is, it cannot be so much further behind that it covers only a small fraction of 
 $\alpha$ Per cluster area, which is quite extended on the sky.
However, $\alpha$ Per cluster members have been found in ground-based studies in an
area on the sky about twice as wide at the PRS ROSAT observations
(Prosser 1992).  Thus a grouping about twice as far away as the $\alpha$ Per
cluster is consistent with the spread of ``bad L$_X$'' stars on the sky.    

In summary, a grouping behind the  $\alpha$ Per cluster is a plausible explanation for the   
``bad L$_X$'' sequence.  This interpretation is key to deriving a XLD
appropriate for the 50 Myr stars in the $\alpha$ Per cluster itself.

\section{Summary} 

We have investigated an archival XMM-Newton image of part of the $\alpha$ Per cluster, partly 
because the age of this cluster makes it appropriate for comparison with low-mass companions
of Cepheids.  In particular, this data adds observations of X-ray faint M stars to 
the wider but shallower observations of the whole cluster area by ROSAT.  The 
count rate sequence for the M stars continues the sequence for hotter dwarfs in both
J magnitude and J--K color.  XLD's have been derived for both the hotter dwarfs, and 
also the M stars. The ``bad L$_X$'' sequence identified by PRS is  
appears to be a 
background grouping at a distance of about 500 pc and an age similar
(or slightly older) than 
the $\alpha$ Per cluster, as shown by near IR magnitudes and colors,
as well as X-ray luminosities and temperatures.

\acknowledgments
IP is grateful to Dr. E. Franciosini for providing an 
unpublished catalog of members of Alpha Persei.  We thank an anonymous
referee for comments that improved the presentation of the paper.
The XMM-Newton guest investigator program supported IP through grant NNX09AP46G.
 Support for this work was also provided  from the Chandra X-ray Center NASA 
Contract NAS8-03060.  
Vizier and  SIMBAD were used in the preparation of this study.

%25  & 	13.308	 & 	12.869	 & 	12.72	  & 	0.18	 & 	$-$	 & 	-14.44	 & 	2T  \\

\clearpage

%\plotone{Eaton_set.eps}

%\plotone{Residuals.eps}

%   \includegraphics[angle=270,width=6.5in]{cudw16.hrd.ps}

\appendix

\begin{table}
\caption{List of X-ray sources detected in EPIC-\xmm\ image.} \small
\begin{tabular}{l l l r r r r} \hline \hline
ID	 & R.A.  & Dec. & 	Offaxis 	 & 	Significance &	Rate & Exp. Time \\ \hline
&     (J2000)	&  (J2000)	& \arcmin	 & 	$\sigma_{bkg}$	&	ct ks$^{-1}$	 & 	ks \\ \hline
1	 & 	03:26:37.9	 & 	48:37:28.9	 & 	15.	 & 	15.7	 & 	5.55	 & 	48.3  \\
2	 & 	03:25:43.7	 & 	48:38:32.4	 & 	14.	 & 	7.76	 & 	1.63	 & 	50.7  \\
3	 & 	03:26:43.6	 & 	48:38:47.7	 & 	14.	 & 	6.75	 & 	1.66	 & 	57.3  \\
4	 & 	03:26:01.3	 & 	48:39:10.1	 & 	12.	 & 	21.7	 & 	4.83	 & 	59.9  \\
5	 & 	03:25:50.9	 & 	48:39:22.2	 & 	13.	 & 	29.4	 & 	7.41	 & 	57.4  \\
6	 & 	03:25:53.0	 & 	48:40:30.2	 & 	11.	 & 	23.6	 & 	5.10	 & 	65.1  \\
7	 & 	03:25:37.7	 & 	48:41:43.5	 & 	11.	 & 	14.8	 & 	2.79	 & 	63.7  \\
8	 & 	03:26:54.5	 & 	48:42:17.0	 & 	12.	 & 	5.31	 & 	0.423	 & 	72.5  \\
9	 & 	03:25:54.7	 & 	48:42:28.4	 & 	9.4	 & 	7.77	 & 	1.48	 & 	78.0  \\
10	 & 	03:26:14.3	 & 	48:42:39.6	 & 	9.0	 & 	33.6	 & 	6.19	 & 	87.3  \\
11	 & 	03:26:59.1	 & 	48:42:45.1	 & 	12.	 & 	60.4	 & 	16.5	 & 	73.9  \\
12	 & 	03:26:24.1	 & 	48:42:46.3	 & 	9.2	 & 	9.78	 & 	1.35	 & 	88.1  \\
13	 & 	03:26:41.3	 & 	48:43:18.9	 & 	10.	 & 	8.97	 & 	1.59	 & 	87.5  \\
14	 & 	03:26:34.0	 & 	48:43:44.8	 & 	9.0	 & 	6.56	 & 	1.05	 & 	92.5  \\
15	 & 	03:26:45.5	 & 	48:43:46.0	 & 	10.	 & 	20.9	 & 	3.34	 & 	77.4  \\
16	 & 	03:26:41.5	 & 	48:43:56.6	 & 	9.5	 & 	5.43	 & 	0.459	 & 	91.3  \\
17	 & 	03:25:29.8	 & 	48:44:00.1	 & 	9.8	 & 	5.89	 & 	1.18	 & 	68.8  \\
18	 & 	03:26:18.4	 & 	48:44:06.8	 & 	7.7	 & 	4.90	 & 	0.396	 & 	98.5  \\
19	 & 	03:26:22.6	 & 	48:44:20.2	 & 	7.7	 & 	5.95	 & 	0.435	 & 	103.  \\
20	 & 	03:26:01.7	 & 	48:44:35.7	 & 	7.1	 & 	11.0	 & 	1.03	 & 	99.9  \\
21	 & 	03:26:12.3	 & 	48:44:37.2	 & 	7.0	 & 	6.29	 & 	0.512	 & 	98.1  \\
22	 & 	03:26:52.7	 & 	48:44:38.3	 & 	10.	 & 	5.96	 & 	0.512	 & 	89.6  \\
23	 & 	03:27:16.4	 & 	48:44:43.2	 & 	13.	 & 	13.0	 & 	2.85	 & 	69.8  \\
24	 & 	03:26:47.2	 & 	48:45:14.9	 & 	9.1	 & 	5.13	 & 	0.562	 & 	96.0  \\
25	 & 	03:25:47.2	 & 	48:45:15.7	 & 	7.2	 & 	6.32	 & 	0.809	 & 	88.4  \\
26	 & 	03:26:41.4	 & 	48:45:23.3	 & 	8.4	 & 	6.94	 & 	0.773	 & 	103.  \\
27	 & 	03:25:56.5	 & 	48:45:40.6	 & 	6.2	 & 	5.34	 & 	0.399	 & 	105.  \\
28	 & 	03:25:13.8	 & 	48:45:56.4	 & 	10.	 & 	5.52	 & 	1.27	 & 	56.2  \\
29	 & 	03:25:27.2	 & 	48:46:03.6	 & 	8.6	 & 	15.1	 & 	1.45	 & 	74.3  \\
30	 & 	03:27:05.5	 & 	48:46:20.4	 & 	11	 & 	11.9	 & 	2.22	 & 	86.8  \\
31	 & 	03:25:26.3	 & 	48:46:28.9	 & 	8.5	 & 	6.38	 & 	0.483	 & 	75.7  \\
32	 & 	03:26:40.8	 & 	48:46:38.7	 & 	7.4	 & 	24.2	 & 	4.84	 & 	55.7  \\
33	 & 	03:25:23.0	 & 	48:46:58.3	 & 	8.7	 & 	10.5	 & 	1.23	 & 	71.4  \\
34	 & 	03:27:02.8	 & 	48:47:07.7	 & 	10.	 & 	33.0	 & 	6.97	 & 	93.4  \\
35	 & 	03:25:32.0	 & 	48:47:24.2	 & 	7.2	 & 	11.0	 & 	1.39	 & 	59.6  \\
36	 & 	03:26:50.2	 & 	48:47:33.3	 & 	8.1	 & 	182.	 & 	63.3	 & 	111.  \\
37	 & 	03:26:53.8	 & 	48:47:34.4	 & 	8.6	 & 	11.2	 & 	0.99	 & 	108.  \\
38	 & 	03:26:11.1	 & 	48:47:45.8	 & 	3.9	 & 	7.64	 & 	0.512	 & 	131.  \\
39	 & 	03:27:09.3	 & 	48:47:48.2	 & 	11.	 & 	9.50	 & 	1.71	 & 	84.7  \\
40	 & 	03:25:35.9	 & 	48:48:05.7	 & 	6.3	 & 	6.23	 & 	0.521	 & 	96.1  \\
41	 & 	03:26:04.2	 & 	48:48:08.5	 & 	3.5	 & 	224.	 & 	78.7	 & 	129.  \\
42	 & 	03:24:58.9	 & 	48:48:12.8	 & 	12.	 & 	94.7	 & 	48.1	 & 	52.7  \\
43	 & 	03:27:16.7	 & 	48:48:39.9	 & 	12.	 & 	6.89	 & 	0.861	 & 	82.1  \\
44	 & 	03:25:54.4	 & 	48:48:49.2	 & 	3.5	 & 	6.36	 & 	0.571	 & 	119.  \\
45	 & 	03:26:32.9	 & 	48:48:52.0	 & 	5.0	 & 	6.41	 & 	0.355	 & 	136.  \\
46	 & 	03:27:10.9	 & 	48:49:00.3	 & 	11.	 & 	9.67	 & 	0.888	 & 	90.4  \\
47	 & 	03:26:30.1	 & 	48:49:22.2	 & 	4.3	 & 	8.51	 & 	0.527	 & 	140  \\
48	 & 	03:25:07.1	 & 	48:49:25.4	 & 	10.	 & 	15.4	 & 	3.26	 & 	52.4  \\
49	 & 	03:27:22.6	 & 	48:50:06.5	 & 	12.	 & 	27.3	 & 	6.26	 & 	77.0  \\
50	 & 	03:25:56.8	 & 	48:50:15.6	 & 	2.2	 & 	5.12	 & 	0.212	 & 	132  \\
51	 & 	03:27:02.5	 & 	48:50:21.4	 & 	9.2	 & 	11.5	 & 	1.44	 & 	79.9  \\
52	 & 	03:26:16.4	 & 	48:50:29.6	 & 	1.8	 & 	117.	 & 	21.3	 & 	149  \\
53	 & 	03:25:40.5	 & 	48:50:41.9	 & 	4.5	 & 	11.8	 & 	2.19	 & 	84.1  \\
54	 & 	03:26:50.3	 & 	48:50:49.6	 & 	7.1	 & 	14.2	 & 	1.73	 & 	103.  \\
55	 & 	03:27:07.8	 & 	48:50:57.2	 & 	10.	 & 	7.97	 & 	1.20	 & 	97.2  \\
56	 & 	03:24:55.6	 & 	48:51:00.1	 & 	12.	 & 	7.16	 & 	1.94	 & 	52.6  \\
57	 & 	03:25:53.4	 & 	48:51:03.5	 & 	2.4	 & 	5.91	 & 	0.512	 & 	128.  \\
58	 & 	03:26:00.8	 & 	48:51:30.9	 & 	1.1	 & 	5.44	 & 	0.506	 & 	88.7  \\
59	 & 	03:25:24.2	 & 	48:51:42.3	 & 	7.1	 & 	8.19	 & 	0.768	 & 	85.2  \\
60	 & 	03:27:02.3	 & 	48:51:55.3	 & 	9.0	 & 	10.0	 & 	1.35	 & 	105.  \\
61	 & 	03:25:19.7	 & 	48:51:57.0	 & 	7.9	 & 	6.87	 & 	0.687	 & 	78.5  \\
62	 & 	03:27:39.6	 & 	48:51:56.7	 & 	15.	 & 	6.78	 & 	1.12	 & 	23.6  \\
63	 & 	03:25:51.0	 & 	48:52:06.0	 & 	2.8	 & 	7.18	 & 	0.65	 & 	124.  \\
64	 & 	03:26:03.3	 & 	48:52:07.9	 & 	0.86	 & 	5.86	 & 	0.586	 & 	139.  \\
65	 & 	03:25:12.4	 & 	48:52:07.6	 & 	9.1	 & 	15.2	 & 	3.56	 & 	70.2  \\
66	 & 	03:27:14.1	 & 	48:52:14.8	 & 	11.	 & 	6.53	 & 	0.559	 & 	86.6  \\
67	 & 	03:26:41.3	 & 	48:52:17.3	 & 	5.6	 & 	16.9	 & 	1.62	 & 	133.  \\
68	 & 	03:24:50.0	 & 	48:52:18.6	 & 	13.	 & 	80.6	 & 	57.8	 & 	36.4  \\
69	 & 	03:26:02.8	 & 	48:52:59.9	 & 	1.6	 & 	5.47	 & 	0.305	 & 	136.  \\
70	 & 	03:24:59.4	 & 	48:53:01.0	 & 	11.	 & 	47.5	 & 	26.1	 & 	32.4  \\
71	 & 	03:27:05.0	 & 	48:53:05.1	 & 	9.6	 & 	8.74	 & 	0.936	 & 	95.5  \\
72	 & 	03:27:07.3	 & 	48:53:19.1	 & 	10.	 & 	15.9	 & 	2.06	 & 	94.9  \\
73	 & 	03:25:31.3	 & 	48:53:21.5	 & 	6.2	 & 	23.1	 & 	3.35	 & 	92.9  \\
74	 & 	03:24:48.3	 & 	48:53:20.9	 & 	13.	 & 	90.6	 & 	85.4	 & 	27.9  \\
75	 & 	03:27:11.3	 & 	48:54:01.5	 & 	11.	 & 	8.36	 & 	0.759	 & 	88.2  \\
76	 & 	03:26:47.4	 & 	48:54:10.7	 & 	7.1	 & 	6.92	 & 	0.637	 & 	119.  \\
77	 & 	03:25:45.9	 & 	48:54:11.8	 & 	4.4	 & 	5.97	 & 	0.541	 & 	111.  \\
78	 & 	03:26:44.1	 & 	48:54:11.8	 & 	6.6	 & 	17.9	 & 	1.90	 & 	108.  \\
79	 & 	03:26:00.9	 & 	48:54:17.4	 & 	2.9	 & 	10.8	 & 	1.26	 & 	128.  \\
80	 & 	03:26:44.1	 & 	48:54:23.5	 & 	6.6	 & 	9.13	 & 	0.848	 & 	111.  \\
81	 & 	03:26:33.2	 & 	48:55:42.2	 & 	5.9	 & 	5.87	 & 	0.488	 & 	123.  \\
82	 & 	03:25:36.2	 & 	48:55:59.1	 & 	6.8	 & 	10.8	 & 	1.38	 & 	90.1  \\
83	 & 	03:25:39.4	 & 	48:56:30.3	 & 	6.7	 & 	5.35	 & 	0.594	 & 	91.4  \\
84	 & 	03:26:31.5	 & 	48:57:20.0	 & 	7.0	 & 	15.2	 & 	1.68	 & 	105.  \\
85	 & 	03:25:28.6	 & 	48:57:53.0	 & 	9.0	 & 	37.8	 & 	8.82	 & 	73.8  \\
86	 & 	03:27:25.0	 & 	48:58:12.5	 & 	14.	 & 	20.6	 & 	5.32	 & 	27.4  \\
87	 & 	03:25:28.7	 & 	48:58:14.3	 & 	9.2	 & 	7.49	 & 	0.862	 & 	72.2  \\
88	 & 	03:25:51.5	 & 	48:58:53.7	 & 	7.8	 & 	29.8	 & 	4.88	 & 	87.4  \\
89	 & 	03:26:55.9	 & 	48:58:57.3	 & 	11.	 & 	20.4	 & 	3.77	 & 	69.8  \\
90	 & 	03:25:10.3	 & 	48:59:44.0	 & 	12.	 & 	20.9	 & 	7.98	 & 	34.9  \\
91	 & 	03:25:39.0	 & 	48:59:57.1	 & 	9.6	 & 	5.37	 & 	0.308	 & 	72.4  \\
92	 & 	03:26:06.0	 & 	49:00:07.0	 & 	8.5	 & 	21.3	 & 	3.24	 & 	86.5  \\
93	 & 	03:26:54.7	 & 	49:01:12.8	 & 	12.	 & 	5.52	 & 	0.545	 & 	70.7  \\
94	 & 	03:26:40.1	 & 	49:01:36.6	 & 	11.	 & 	5.46	 & 	0.522	 & 	74.7  \\
95	 & 	03:25:51.5	 & 	49:02:07.6	 & 	11.	 & 	7.80	 & 	1.43	 & 	45.1  \\
96	 & 	03:26:27.7	 & 	49:02:13.5	 & 	11.	 & 	101.	 & 	35.1	 & 	71.9  \\
97	 & 	03:25:48.2	 & 	49:02:40.4	 & 	12.	 & 	5.74	 & 	0.625	 & 	23.9  \\
98	 & 	03:26:39.8	 & 	49:02:47.8	 & 	12.	 & 	14.3	 & 	3.37	 & 	66.9  \\
99	 & 	03:26:32.2	 & 	49:03:13.5	 & 	12.	 & 	24.6	 & 	5.31	 & 	60.8  \\
100	 & 	03:26:23.4	 & 	49:04:16.4	 & 	13.	 & 	6.47	 & 	1.60	 & 	13.9  \\
101	 & 	03:26:01.9	 & 	49:04:42.3	 & 	13.	 & 	14.9	 & 	5.41	 & 	22.9  \\
102	 & 	03:26:26.3	 & 	49:04:54.7	 & 	14.	 & 	6.26	 & 	3.12	 & 	12.9  \\ \hline
\end{tabular}
\end{table}

\begin{table}
\caption{List of sources with 2MASS match and parameters from best fit modeling of X-ray spectra.}
\begin{tabular}{llllrrrllll}\\\hline\hline \tiny
 ID	&	R.A. 	 &	Dec. 	 & 	2MASS ID		&  J	 & 	H   & 	K  & 	kT	 &	E.M.	 &	$\log f_X$	 & 	$\log L_X$	\\ 
      &      (J2000)	 & 	(J2000)	& 				& mag		& 	mag	 & mag	&  keV	 &	cm$^{-3}$	& 	erg s$^{-1}$ cm$^{-2}$& 	erg s$^{-1}$	\\\hline
4	 & 	03:26:01.3	 & 	48:39:09.7	 & 	03260131+4839097	 & 	12.60	 & 	11.92	 & 	11.69	 & 	0.76	 & 	51.73	 & 	-13.8	 & 	28.8  \\
10	 & 	03:26:14.2	 & 	48:42:38.3	 & 	03261419+4842382	 & 	12.48	 & 	11.80	 & 	11.55	 & 	0.89	 & 	51.92	 & 	-13.6	 & 	29.0  \\
19	 & 	03:26:22.7	 & 	48:44:20.1	 & 	03262270+4844201	 & 	12.77	 & 	12.31	 & 	12.12	 & 	0.86	 & 	50.65	 & 	-14.9	 & 	27.7  \\
22	 & 	03:26:52.6	 & 	48:44:37.9	 & 	03265263+4844378	 & 	11.10	 & 	10.77	 & 	10.76	 & 	0.88	 & 	51.06	 & 	-14.4	 & 	28.1  \\
42	 & 	03:24:58.8	 & 	48:48:14.7	 & 	03245884+4848147	 & 	16.47	 & 	15.69	 & 	14.52	 & 	--	 & 	--	 & 	--	 & 	--  \\
33	 & 	03:25:23.0	 & 	48:46:56.2	 & 	03252295+4846562	 & 	13.64	 & 	13.04	 & 	12.78	 & 	0.81	 & 	51.24	 & 	-14.3	 & 	28.2  \\
35	 & 	03:25:31.8	 & 	48:47:21.4	 & 	03253182+4847213	 & 	13.80	 & 	13.18	 & 	12.98	 & 	0.81	 & 	51.38	 & 	-14.1	 & 	28.4  \\
25	 & 	03:25:47.2	 & 	48:45:13.6	 & 	03254715+4845136	 & 	13.31	 & 	12.87	 & 	12.72	 & 	0.18	 & 	53.95	 & 	-11.9	 & 	30.7  \\
27	 & 	03:25:56.6	 & 	48:45:38.1	 & 	03255664+4845381	 & 	15.65	 & 	14.90	 & 	14.67	 & 	--	 & 	--	 & 	--	 & 	--  \\
41	 & 	03:26:04.2	 & 	48:48:07.1	 & 	03260421+4848070	 & 	9.49	 & 	9.21	 & 	9.16	 & 	0.64	 & 	53.17	 & 	-12.4	 & 	30.2  \\
38	 & 	03:26:11.1	 & 	48:47:45.0	 & 	03261110+4847450	 & 	15.02	 & 	14.37	 & 	14.04	 & 	--	 & 	--	 & 	--	 & 	--  \\
32	 & 	03:26:40.8	 & 	48:46:36.8	 & 	03264075+4846368	 & 	8.76	 & 	8.58	 & 	8.57	 & 	0.47	 & 	52.18	 & 	-13.4	 & 	29.1  \\
26	 & 	03:26:41.2	 & 	48:45:21.0	 & 	03264121+4845210	 & 	14.6	 & 	13.86	 & 	13.67	 & 	--	 & 	--	 & 	--	 & 	--  \\
36	 & 	03:26:50.1	 & 	48:47:32.1	 & 	03265010+4847320	 & 	9.20	 & 	9.01	 & 	8.89	 & 	0.64	 & 	53.10	 & 	-12.4	 & 	30.1  \\
39	 & 	03:27:09.4	 & 	48:47:49.2	 & 	03270936+4847491	 & 	12.03	 & 	11.63	 & 	11.56	 & 	0.53	 & 	52.62	 & 	-13.1	 & 	29.4  \\
30	 & 	03:27:05.7	 & 	48:46:20.1	 & 	03270565+4846201	 & 	16.70	 & 	16.05	 & 	16.33	 & 	--	 & 	--	 & 	--	 & 	--  \\
68	 & 	03:24:49.7	 & 	48:52:18.4	 & 	03244971+4852183	 & 	10.11	 & 	9.72	 & 	9.59	 & 	0.90	 & 	52.89	 & 	-12.6	 & 	29.9  \\
48	 & 	03:25:07.2	 & 	48:49:26.0	 & 	03250721+4849260	 & 	13.12	 & 	12.4	 & 	12.20	 & 	--	 & 	--	 & 	--	 & 	--  \\
65	 & 	03:25:12.3	 & 	48:52:05.3	 & 	03251232+4852052	 & 	9.77	 & 	9.56	 & 	9.47	 & 	0.53	 & 	51.69	 & 	-13.9	 & 	28.7  \\
59	 & 	03:25:24.3	 & 	48:51:40.8	 & 	03252428+4851407	 & 	12.38	 & 	12.02	 & 	11.92	 & 	0.34	 & 	52.07	 & 	-13.6	 & 	28.9  \\
63	 & 	03:25:50.7	 & 	48:52:05.3	 & 	03255074+4852052	 & 	13.20	 & 	12.72	 & 	12.59	 & 	0.54	 & 	51.69	 & 	-13.9	 & 	28.7  \\
58	 & 	03:26:01.0	 & 	48:51:30.9	 & 	03260103+4851309	 & 	15.37	 & 	14.97	 & 	14.59	 & 	--	 & 	--	 & 	--	 & 	--  \\
47	 & 	03:26:30.0	 & 	48:49:20.9	 & 	03262999+4849209	 & 	14.08	 & 	13.49	 & 	13.22	 & 	0.73	 & 	50.92	 & 	-14.6	 & 	27.9  \\
52	 & 	03:26:16.4	 & 	48:50:28.4	 & 	03261639+4850284	 & 	10.81	 & 	10.32	 & 	10.21	 & 	0.77	 & 	52.45	 & 	-13.1	 & 	29.5  \\
67	 & 	03:26:41.3	 & 	48:52:16.1	 & 	03264130+4852160	 & 	11.83	 & 	11.54	 & 	11.44	 & 	0.65	 & 	51.71	 & 	-13.8	 & 	28.7  \\
51	 & 	03:27:02.4	 & 	48:50:19.4	 & 	03270241+4850193	 & 	13.18	 & 	12.56	 & 	12.30	 & 	0.58	 & 	51.14	 & 	-14.4	 & 	28.1  \\
62	 & 	03:27:39.6	 & 	48:51:54.0	 & 	03273962+4851539	 & 	12.13	 & 	11.72	 & 	11.60	 & 	0.60	 & 	51.49	 & 	-14.0	 & 	28.5  \\
74	 & 	03:24:48.4	 & 	48:53:19.9	 & 	03244838+4853199	 & 	10.08	 & 	9.55	 & 	9.42	 & 	0.96	 & 	53.06	 & 	-12.5	 & 	30.1  \\
82	 & 	03:25:36.1	 & 	48:55:57.4	 & 	03253612+4855573	 & 	13.52	 & 	12.93	 & 	12.57	 & 	0.82	 & 	51.32	 & 	-14.2	 & 	28.4  \\
77	 & 	03:25:46.0	 & 	48:54:10.0	 & 	03254597+4854100	 & 	15.95	 & 	15.21	 & 	14.99	 & 	--	 & 	--	 & 	--	 & 	--  \\
69	 & 	03:26:02.8	 & 	48:52:59.2	 & 	03260275+4852591	 & 	16.26	 & 	15.64	 & 	15.12	 & 	--	 & 	--	 & 	--	 & 	--  \\
71	 & 	03:27:05.1	 & 	48:53:04.5	 & 	03270505+4853044	 & 	10.57	 & 	9.981	 & 	9.78	 & 	0.97	 & 	51.40	 & 	-12.7	 & 	29.8  \\
75	 & 	03:27:11.2	 & 	48:53:60.0	 & 	03271116+4853599	 & 	14.62	 & 	13.87	 & 	13.67	 & 	--	 & 	--	 & 	--	 & 	--  \\
90	 & 	03:25:10.4	 & 	48:59:45.4	 & 	03251041+4859453	 & 	12.59	 & 	12.01	 & 	11.75	 & 	1.01	 & 	52.25	 & 	-13.3	 & 	29.3  \\
85	 & 	03:25:28.7	 & 	48:57:51.5	 & 	03252866+4857515	 & 	11.43	 & 	10.98	 & 	10.84	 & 	0.96	 & 	52.11	 & 	-13.4	 & 	29.1  \\
87	 & 	03:25:28.9	 & 	48:58:14.3	 & 	03252893+4858143	 & 	14.53	 & 	13.99	 & 	13.88	 & 	--	 & 	--	 & 	--	 & 	--  \\
86	 & 	03:27:24.9	 & 	48:58:11.9	 & 	03272491+4858118	 & 	13.53	 & 	12.88	 & 	12.62	 & 	0.97	 & 	52.03	 & 	-13.5	 & 	29.0  \\
92	 & 	03:26:06.0	 & 	49:00:06.4	 & 	03260595+4900064	 & 	15.64	 & 	14.79	 & 	14.44	 & 	34	 & 	52.58	 & 	-12.8	 & 	29.8  \\
96	 & 	03:26:27.6	 & 	49:02:12.4	 & 	03262764+4902123	 & 	10.77	 & 	10.26	 & 	10.11	 & 	0.94	 & 	52.75	 & 	-12.8	 & 	29.8  \\ \hline
\end{tabular}
\end{table}

\end{document}